\documentclass{amsproc}
\usepackage{amsmath,amssymb,amsthm,amsfonts}

\usepackage{bbm}

\usepackage[colorlinks=true,citecolor=blue]{hyperref}

\usepackage{graphicx}

\numberwithin{equation}{section}

\theoremstyle{plain}

\newtheorem*{theorem*}{Theorem}

\theoremstyle{remark}

\begin{document}

\title[Algorithmic Uncertainty]{Uncertainty in Criminal Justice Algorithms:  simulation studies of the Pennsylvania Additive Classification Tool}

\author[S. Dhar]{Swarup Dhar}
\address{Department of Mathematics, Bucknell University}
\author[V. Massaro]{Vanessa Massaro}
\address{Department of Geography, Bucknell University}
\author[D. Mir]{Darakhshan Mir}
\address{Department of Computer Science, Bucknell University}
\author[N.C. Ryan]{Nathan C. Ryan}
\address{Department of Mathematics, Bucknell University}


\maketitle

\begin{abstract}
Much attention has been paid to algorithms related to sentencing, the setting of bail, parole decisions and recidivism while less attention has been paid to carceral algorithms, those algorithms used to determine an incarcerated individual's lived experience.  In this paper we study one such algorithm, the Pennsylvania Additive Classification Tool (PACT) that assigns custody levels to incarcerated individuals.  We analyze the PACT in ways that criminal justice algorithms are often analyzed: namely, we train an accurate machine learning model for the PACT; we study its fairness across sex, age and race; and we determine which features are most important.  In addition to these conventional computations, we propose and carry out some new ways to study such algorithms.  Instead of focusing on the outcomes themselves, we propose shifting our attention to the variability in the outcomes, especially because many carceral algorithms are used repeatedly and there can be a propagation of uncertainty.  By carrying out several simulations of assigning custody levels, we shine light on problematic aspects of tools like the PACT.
\end{abstract}
 
\section{Introduction}

As has been well-established, the use of algorithms in decision making, low- and high-stakes decisions alike, is pervasive in industry and, increasingly, in government.  Particular domains where decisions are made using algorithms include online advertising, lending and banking, pretrial detention, to name a few.  These decisions are based on predictions which are themselves based on data.  A great deal has been written on biases that are manifest in these processes:  biases that emerge from sampling issues and measurement error and biases in outcomes.  There have been many articles written on various approaches to measuring the bias and, conversely, the fairness of decision making processes in which, often, the details of the algorithm or the predictions that undergird the decision are unknown.  See \cite{mitchell} for an excellent and insightful review of the various ways bias and fairness have been measured and interpreted, as well as measures of fairness based only on data, and those based on various kinds of models.  

An aspect of many of these decision-making algorithms is that they are often reapplied to the same individual at various times.  Lum and Isaac \cite{lum}, describe the impact of training a predictive policing algorithm on biased data.  They describe a feedback loop in predictive policing of drug crimes: in particular, they observe that using algorithms to determine where to police, results in over-policed communities becoming even more disproportionately over-policed. Ensign, \textit{et al} \cite{ensign} have used Polya urn models to give a mathematical explanation for these feedback loops.  

Studying feedback loops is one way to get a handle on the variability in decisions that is generated with the reapplication of predictive algorithms.  Another approach to understanding the variability in decisions is to use sensitivity analysis and simulation.  Blumstein \cite{blumstein} used this approach to model complete criminal justice systems  and to understand which variables most influenced the cost and the flow of offenders through the system.  A more recent study can be found in \cite{dabbaghian}.  Moranian \textit{et al} \cite{moranian} carried out a simulation study for juvenile courts to determine which variables influenced the rate of flow most strongly.  

In addition to these two particular approaches, there has recently been a large number of simulation studies of the criminal justice system.  For example, Cort\'{e}s and Ghosh propose an agent based model in \cite{cortes}. A recent book by Liu and Eck \cite{liu} is dedicated to crime simulation using GIS and various mathematical simulation methods.

In this paper, we carry out a number of simulations to understand a particular tool used by the Pennsylvania Department of Corrections (PADOC).  This tool is called the Pennsylvania Additive Classification Tool (PACT) and is used by the PADOC to assign a custody level to an incarcerated person.  The authors have recently carried out a historical and meta-analysis of the data used by this tool; see \cite{dmmr1}.  There we point out that custody-level determination by PACT is biased by race, that the data is problematic both in how much is missing and in how highly its quality is deemed to be despite it containing errors; that the algorithm uses biased input;  and that the tool itself has competing goals as it is both supposed to reflect both the incarcerated person's rehabilitation and their securitization by the PADOC.

\begin{figure}
\centering
\includegraphics[width=\textwidth]{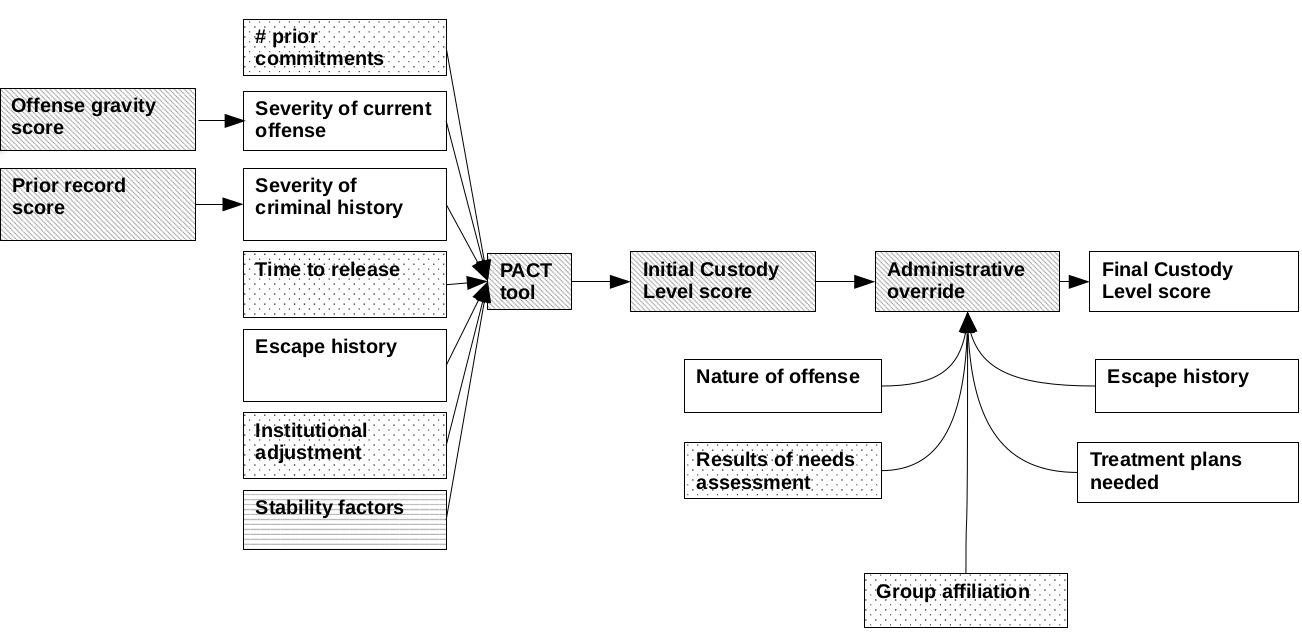}
\caption{A schematic description of how an incarcerated person's custody level score is determined.  The process has two main parts:  a score determined by the PACT and then an override.  Variables in boxes with a background of diagonal lines indicate a lack of transparency about how that variable is calculated; variables in boxes with a background made of little dots are ones where there the values are known to have a societal bias; the variable in a box whose back ground is horizontal lines represents several variables (e.g., marital status, employment status, etc.) that are somewhat arbitrary; and variables in white are ones that are either not known to be biased or whose method of calculation is known. }\label{fig:flowchart}
\end{figure}

Figure~\ref{fig:flowchart} summarizes the data and the steps that go into the PACT tool.  Our interest is in the certainty one can reasonably have in the output of such a flawed tool; we provide a number of methods to quantify or describe the uncertainty one should reasonably hold about the tool's output.  

In addition to these findings about this particular tool, we propose a number of ways to study and understand the uncertainty of any algorithm whose inner workings are obscured to the public.  While much work has been done on bias in outcomes and fairness in decisions, we propose that studying the uncertainty and variability of a model can provide different insights about its impact.

The paper is organized as follows.  In Section~\ref{sec:pact} we describe the PACT in more detail and the data set we are working with.  In Section~\ref{sec:sa} we train a random forest model on our data and then use it to carry out a number of simulation experiments.  In Section~\ref{sec:cf} we carry out fairness analyses, including a measure built around counterfactuals which is, in a sense, a combination of the philosophy behind the previous two sections.  We conclude with a discussion and description of future directions.

\section{The PACT}\label{sec:pact}

The PACT is an example of a \emph{carceral} algorithm; that is, an algorithm that determines an incarcerated person's experience during incarceration.  Most algorithms studied in the context of criminal justice are related to parole, bail and sentencing and studying a carceral algorithm like the PACT presents a number of challenges, including gaining access to the data.  The PACT was introduced by the PADOC in 1991 and generates a raw score that determines an incarcerated person's custody level, ranging from a custody level of 1 (community corrections) up to 5 (maximum security).  The Pennsylvania state documents (see \cite{PACT}) related to PACT tell us that the algorithm ``is confidential and not for public dissemination'' and so analyzing it fully is impossible.  From these same documents we were able to develop the flowchart in Figure~\ref{fig:flowchart}.  

The assignment of custody levels has four main steps.  First, the PACT tool is applied to data transmitted to the PADOC and an initial score is derived and then that score is turned into a number from 1 to 5, representing that person's custody level.  At that point, the particular prison can decide to override this score either for administrative reasons (e.g., number of beds) or for other reasons at the discretion of the particular prison.  At the end of each year a similar process of reclassification is carried out.  A reclassification score is determined algorithmically and then, once again, the prison can decide to override that score for either administrative or discretionary reasons.  

\subsection{Summary of the data}

\begin{figure}
\centering
\includegraphics[width=\textwidth]{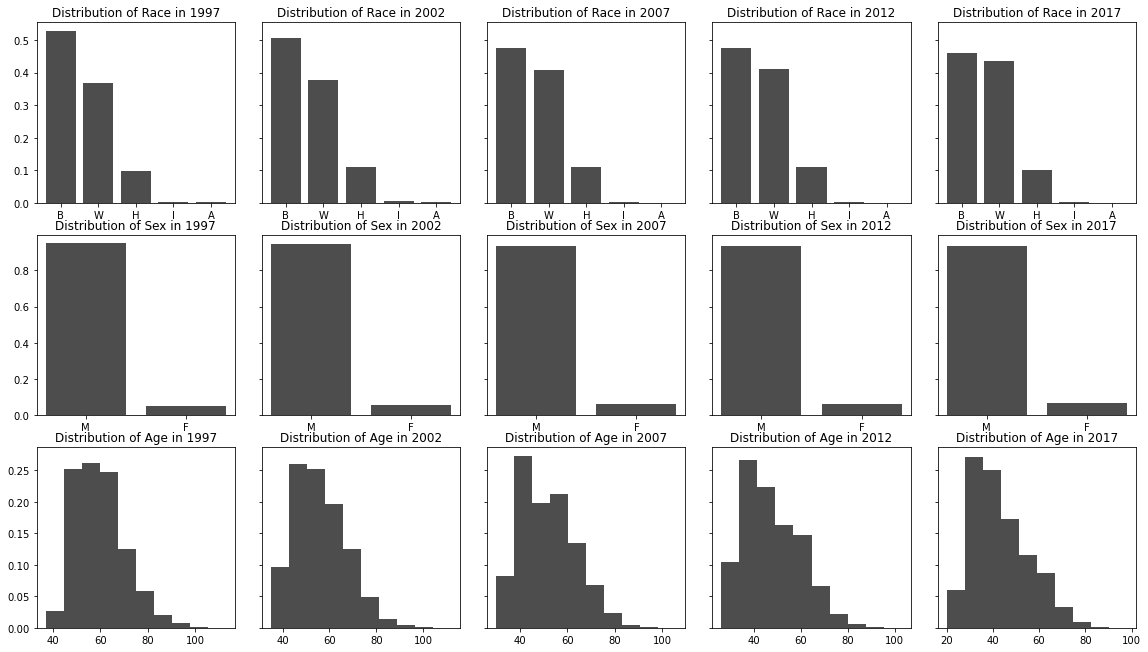}
\caption{Basic demographic data.  In the first row, we see the distribution of the race or ethnicity of the incarcerated people in 1007, 2002, 2007, 2012 and 2017.  Here B means Black, W means White, H means Hispanic, I means American Indian, and A means Asian and Pacific Islander.  The second row gives the distribution of the sex of incarcerated people in the same years.  The third row gives the distribution of the ages.}\label{fig:demographics}
\end{figure}

In July 2018 we requested a data pull from the PADOC using the Pennsylvania Department of Corrections Research Approach Request Form (RARF).  The intent of the request was to study factors that influence parole decisions. The PACT is one such factor and, because of its importance, we decided to study it more closely.  We requested data on incarcerated people, including those who have been paroled, who were in the system in 1997, 2002, 2007, 2012, and 2017.  We requested variables that were related to parole decisions.  Nine months after our initial request we received access to data on more than 280,000 distinct incarcerated people; of those only 146,793 were incarcerated in the years we requested.  See Figure~\ref{fig:demographics} for a graph of the distributions of the demographic variables of the people in our data set.

\subsection{Descriptions of variables}

Most of the variables that we used in our models (see Table~\ref{tbl:rfs}) are self-explanatory and map unambiguously on to the steps laid out in Figure~\ref{fig:flowchart}.  There are some that merit more explanation.  

In principle (as opposed to in the data), every incarcerated person should have a prior record score and the crimes for which they have been most recently committed should have an offense gravity score.  A prior record score is a number from 1 to 4 that indicates a person's criminal history and the number is identified by a statute number in the Pennsylvania Criminal Code.  In the data that we have, we are given the statute number but not the person's prior record score and each statute number is often associated to a range of numbers.  An offense gravity score is similar but ranges from 1 to 15 and indicates the severity and nature of the crime for which the person has been committed.  Like prior record scores, offense gravity scores are not given in the data but a statute is given and the statute, again, has a range of scores associated to it.  In both of these cases, we use the maximum scores associated to a statute, as described in Table~\ref{tbl:rfs}.

Also in Table~\ref{tbl:rfs}, we take two different approaches to reporting a person's age.  For initial classification we use binary variables to describe a person's age but in reclassification we use the numerical age.  This is because of how we trained the reclassification models: in our simulations we apply these models yearly and so having a finer handle on a person's age is necessary in order to have as good a model as possible.

Finally, a difference between reclassification and initial classification is that in initial classification a person's institutional adjustment is determined in an opaque way.  There is no direct analogue for reclassification and so we use the number of disciplinary reports a person has received, instead.  While the number of disciplinary reports is not opaque, it has been shown, in some cases, that the writing of such reports is biased (see, for example, \cite{nytimes}).

\section{Simulation experiments and sensitivity analysis}\label{sec:sa}

In this section we describe a number of simulation experiments that capture the uncertainty of the PACT procedure.  We model the PACT procedure without including the override processes for several reasons.  The main practical reason is there is a lot of missing data and if we limit our model to those incarcerated people for which we have all the necessary decisions (initial classification and an override and reclassification and an override), there would only be a handful of people left to build and test the model on.  A philosophical reason is that we are interested in the workings of the algorithm and not the incidental and subjective contribution that arises in the override process.

\subsection{Random forest models for initial classification and reclassification}

In our experiments, we use random forests to model both the initial classification and reclassification.  Our code is written in Python and uses Scikit-learn \cite{scikit-learn}; our code can be found at \cite{repo}.  The variables we use to train the models are described in Table~\ref{tbl:rfs}.  

The importance of each variable (as measured by mean decrease in impurity) is also listed in Table~\ref{tbl:rfs}.  We observe that for initial classification the most important variables are, in order,  the person's institutional adjustment, the gravity score of the offense committed by the person, the number of previous times the person has been committed to the PADOC and the person's prior record score.  We observe that a person's race, age and marital status also appear to be important.  For reclassification, we find that the number of disciplinary reports a person gets is far and away the most important, with the person's age being the second most important.  After these two variables, the prior record score, previous commitments, gravity score are the next most important.

The random forest model for initial classification has an accuracy of $0.79$ and the one for reclassification has an accuracy of $0.78$.  

\begin{table}\tiny
\begin{tabular}{|p{.7in}||p{2.6in}|p{.45in}|p{.45in}|}\hline
Variable & Description & IC & RE\\\hline
gender\_female & A binary variable indicating whether the person is identified as female (1) or male (0)&	0.010365&	0.005438\\\hline
age\_gt\_45& A binary variable indicating whether the person is older than 45 years old (1) or not  (0)	& 0.046948	& \\\hline
age\_lt\_25& A binary variable indicating whether the person is younger than 25 years old (1) or not  (0)	&	0.026485 & \\	\hline
age	& A quantitative variable used in the reclassification model & & 	0.188189\\\hline
race\_B	&  A binary variable indicating whether the person is identified as Black (1) or not (0)& 0.029642&	0.016741\\\hline
race\_A	&  A binary variable indicating whether the person is identified as Asian of Pacific Islander (1) or not (0)& 0.000863&	0.000351\\\hline
race\_H	&  A binary variable indicating whether the person is identified as Hispanic (1) or not (0)& 0.014115&	0.011085\\\hline
race\_I	&  A binary variable indicating whether the person is identified as American Indian (1) or not (0)& 0.000223&	0.000403\\\hline
race\_O	&  A binary variable indicating whether the person is identified as belonging to a Nonwhite race or ethnicity other than Black, Asian, Hispanic or American Indian (1) or not (0)& 0.001427&	0.001103\\\hline
off\_1\_prs\_max& A quantitative variable representing a person's maximum prior record score &	0.083801	&0.062925\\\hline
off\_1\_gs\_max& A quantitative variable representing the maximum gravity score a person could have received &	0.203287&	0.089114\\\hline
ic\_custdy\_level & A person's initial custody level upon entering the carceral system & &		0.038942\\\hline
prior\_commits & The number of times the person has been previously committed to the PADOC &	0.148917&	0.10117\\\hline
ic\_institut\_adj & A person's institutional adjustment score used during initial classification &	0.235557&\\	\hline
re\_discip\_reports& The number of disciplinary reports a person was given since the previous reclassification &&		0.434796\\\hline
escape\_hist\_1&	 A binary variable representing whether there was an attempted escape &0.018138&	0.010878\\\hline
escape\_hist\_2&	A binary variable representing whether there was a second attempted escape &0.015903&	0.010201\\\hline
escape\_hist\_3&	A binary variable representing whether there was a third attempted escape&0.014793&	0.004418\\\hline
escape\_hist\_4&	A binary variable representing whether there was a fourth attempted escape &0.015923&	0.013931\\\hline
escape\_hist\_5&	A binary variable representing whether there was a fifth attempted escape&0.008083&	0.010316\\\hline
mrt\_stat\_DIV & A binary variable representing whether the person was divorced at the time of commitment (1) or not (0)	&0.033823 &\\	\hline
mrt\_stat\_SEP& A binary variable representing whether the person was separated at the time of commitment (1) or not (0)	&0.020496&\\\hline	
mrt\_stat\_MAR& A binary variable representing whether the person was married at the time of commitment (1) or not (0)	&0.028329&\\	\hline
mrt\_stat\_WID& A binary variable representing whether the person was widowed at the time of commitment (1) or not (0)	&0.006492&\\	\hline
employed & A binary variable representing whether the person was employed at the time of commitment (1) or not (0)	&0.036389	&\\\hline
\end{tabular}
\caption{A summary of the variables used in the construction of the random forest.  The third and fourth columns are, respectively, the importance (as measured by mean decrease in impurity) of the variable in the random forest models we use for initial classification and reclassification; taken together these importance scores also indicate which variables were used in which model.}\label{tbl:rfs}
\end{table}

\subsection{Experiments}

In this section we describe the experiments we carry out.  The goal of these experiments is to explore what happens when a new observation is pushed through the initial classification model.  These new observations are very similar to people in the data set but have been changed in particular ways. While we have also trained a model for the reclassification process, in most of this section we report only on the results for the initial classification of people.  We do this for two reasons:  (1) when we start subsetting the data, we end up with too little data on which to train a reclassification model and (2) we want to focus on the uncertainty at the very beginning of the process. 

\subsubsection{Experiment 1} In this first experiment, we take all people at a fixed custody level, and construct a new sample space.  This new sample space consists of the same variables but the values for each variable come from the multiset of the people in the fixed custody level.  Then, to sample from this new sample space, we randomly sample a value for each variable from each multiset.  An observation generated in this way has values for each variable that someone else in the same custody level has and so, in this sense, this new observation could have been in this data set but differs from people in the data in some number of variables.

\begin{figure}
\centering
\includegraphics[width=\textwidth]{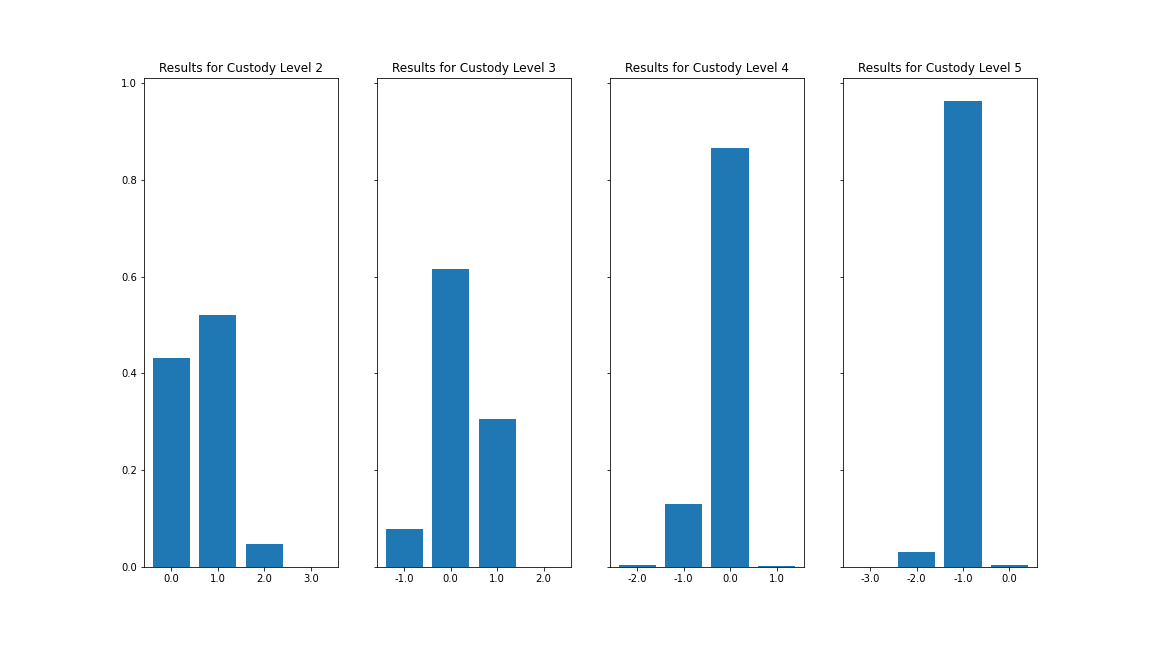}

\caption{Results of Experiment 1 for initial classification.  The bar charts report the frequency with which a synthetic observation in a particular custody level changes to a different custody level and by how many custody levels they would change (e.g., -3.0 means the synthetic observation is classified by our random forest model to three custody levels lower than the one the synthetic observation was generated from).}\label{fig:emp-by-CL}
\end{figure}

In Figure~\ref{fig:emp-by-CL} we report changes in initial custody level for 100 new observations in each custody level, generated as described above. We observe that a fairly large fraction of synthetic observations in each custody level would end up in a different custody level when classified with the model we trained at the beginning.  

\subsubsection{Experiment 2} This experiment is very much like Experiment 1, where, again, we fix a custody level.  In this experiment we generate a synthetic observation by taking a person in this custody level, randomly choosing a single variable and then choosing a new value for that variable from the multiset of values of all people at that custody level.  The synthetic observations, then, will be different from a person in the data in exactly one randomly chosen column.  This experiment is more controlled than Experiment 1.

\begin{figure}
\centering
\includegraphics[width=\textwidth]{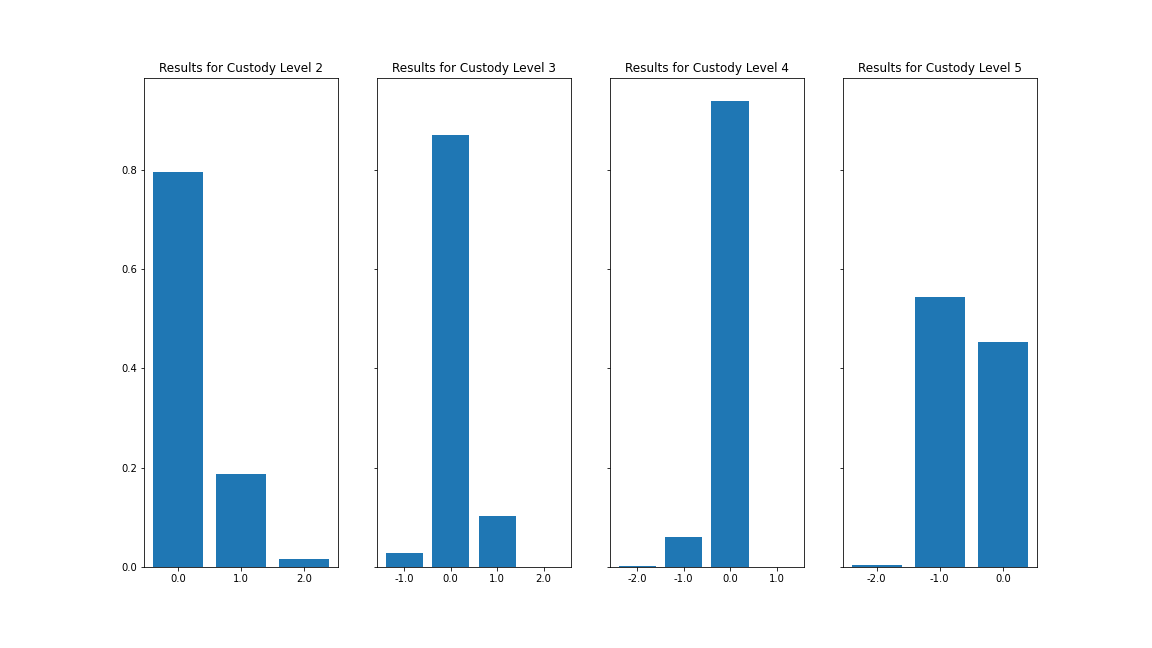}
\\
\includegraphics[width=\textwidth]{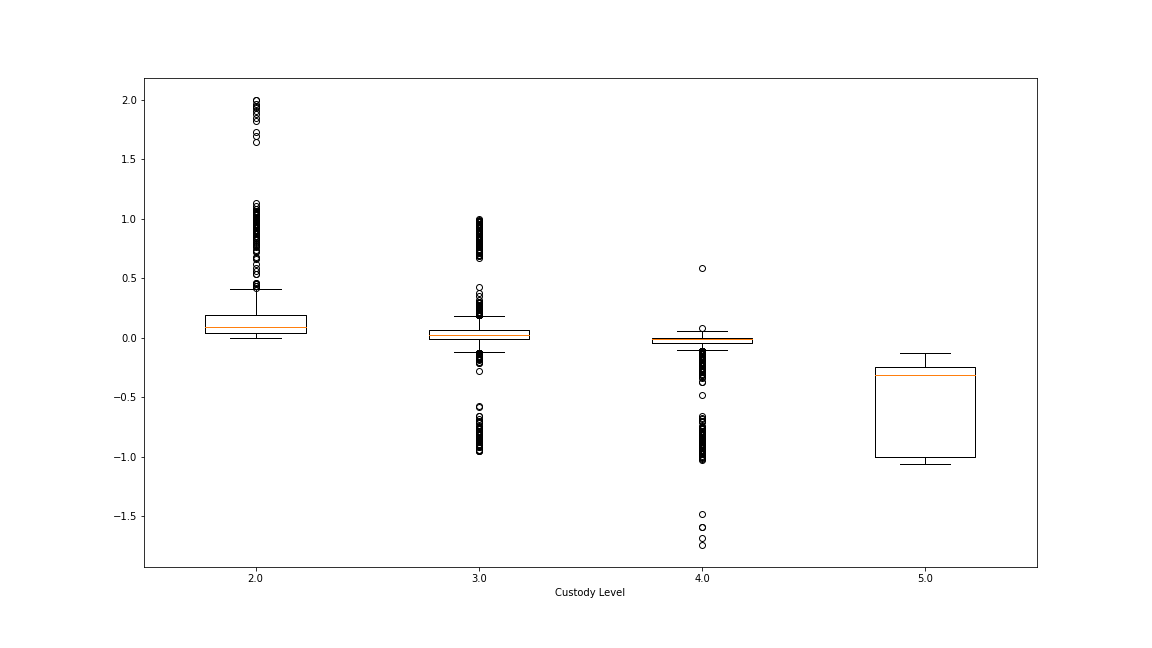}
\caption{Results of Experiment 2 for initial classification.  The bar chart reports the frequency with which a synthetic observation in a particular custody level changes to a different custody level and by how many custody levels they would change (e.g., $-2.0$ means the synthetic observation is classified by our random forest model to three custody levels lower than the one the synthetic observation was generated from).  The box plot shows the new custody after changing a single variable for a person and then applying the random forest model for initial classification.}\label{fig:one-col-by-CL}
\end{figure}

In Figure~\ref{fig:one-col-by-CL} we observe that in initial classification, there is a lot of variability in how a small change to a person in custody level 5 changes their custody level.  For custody levels 2, 3 and 4, there is less variability but there is still a significant number of people for whom a small tweak would send them to a different custody level.

\subsubsection{Experiment 3}  In this third experiment, we proceed as in Experiment 1, but now first stratify for a person's race and then generate the multisets for each race.  In particular, we examine the changes in custody level for people identified as being Black and for those identified as being White.

\begin{figure}
\centering
\includegraphics[width=\textwidth]{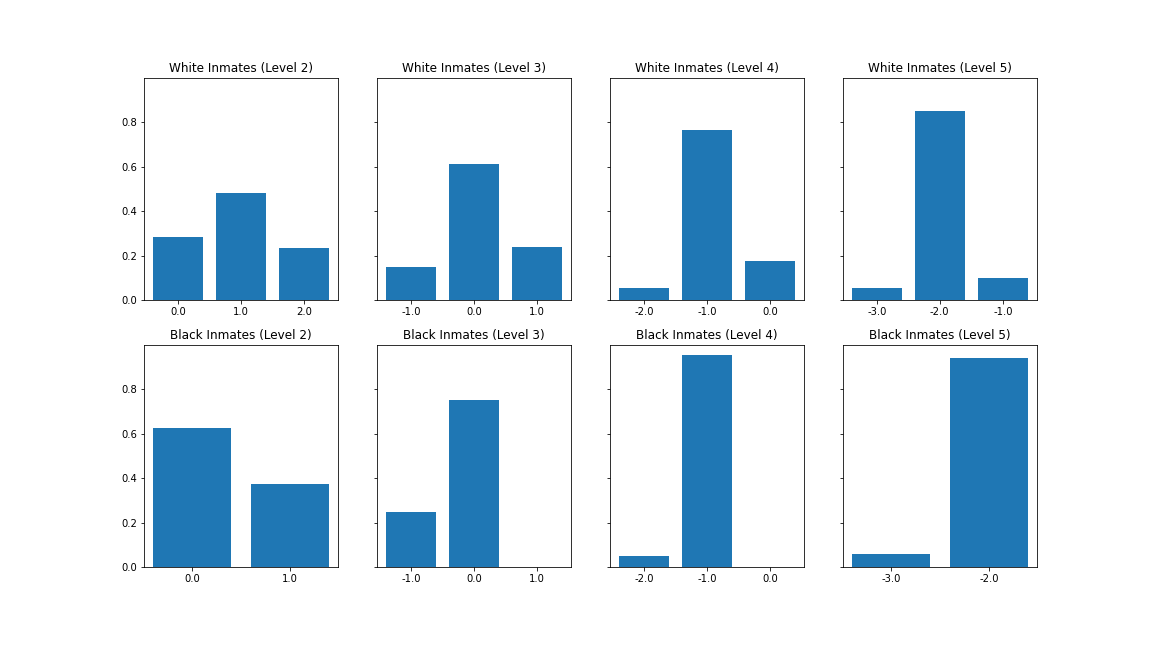}

\caption{Results of Experiment 3 for initial classification for the Black population (bottom) and for the White population (top).  The bar charts report the frequency with which a synthetic observation of a particular race and in a particular custody level changes to a different custody level and by how many custody levels they would change (e.g., $-3.0$ means the synthetic observation is classified by our random forest model to three custody levels lower than the one the synthetic observation was generated from).}\label{fig:emp-by-CL-and-race}
\end{figure}

In Figure~\ref{fig:emp-by-CL-and-race}, we observe the following.  If the person is Black and in a lower custody level, they are, in this experiment, more likely to be classified to a lower level or to be kept at the same level.  On the other hand, if the person is White, they are, in this experiment, more likely to be classified higher or to be kept at the same custody level.  Conversely, if the person is at a high custody level, they will be, at least in the context of this experiment, classified to a lower level than they are to be kept at the same level or classified to a higher level.  The proportion of Black inmates who would be classified in this experiment to a custody level of 2 or 3 is greater than it is for White inmates.

\subsubsection{Experiment 4} In this fourth experiment, we proceed as in Experiment 2, but, like in Experiment 3, we now stratify according to a person's race.  In particular, we examine the changes in custody level for people identified as Black and for people identified as White.  

\begin{figure}
\centering
\includegraphics[width=\textwidth]{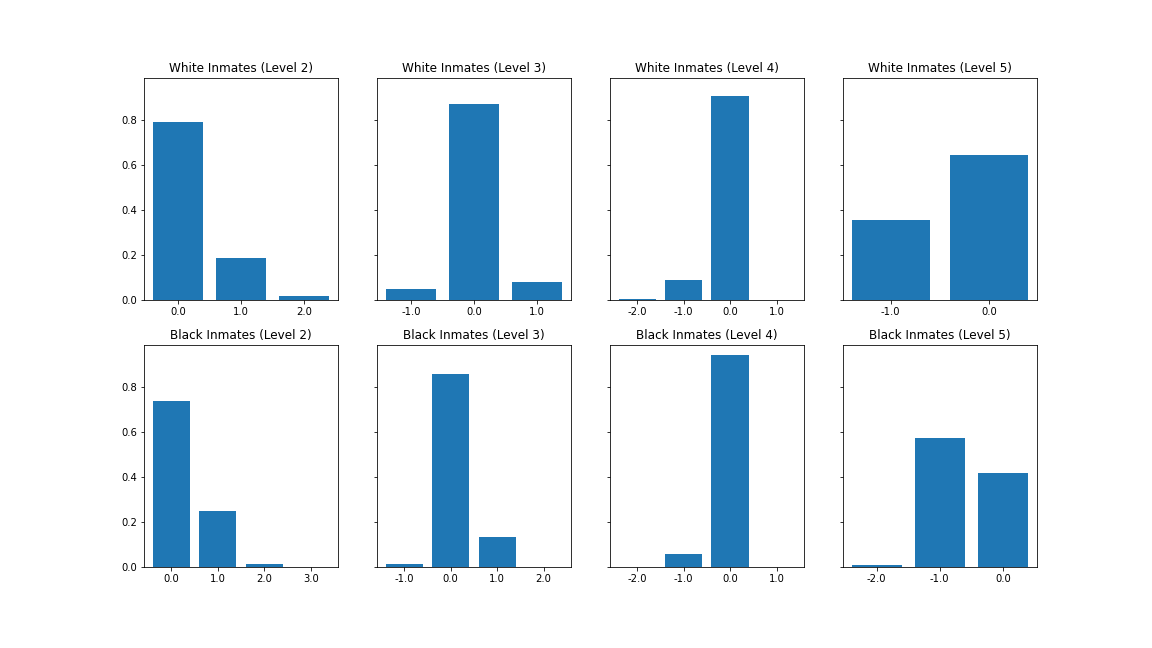}
\\
\includegraphics[width=\textwidth]{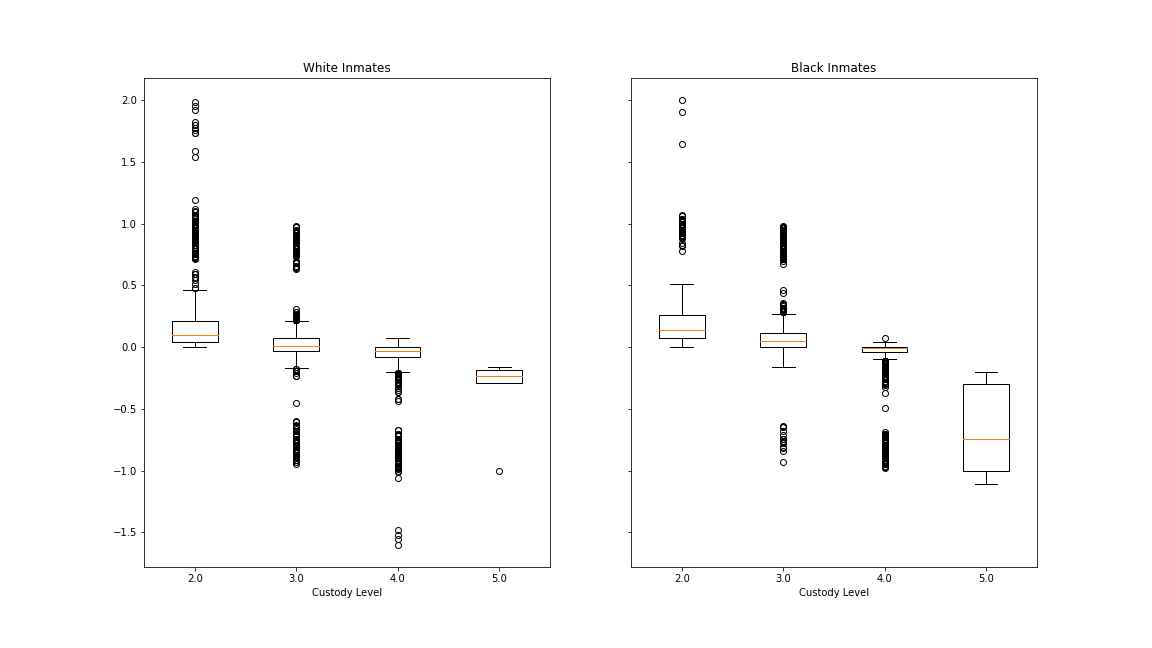}
\caption{Results of Experiment 4 for the Black population and the White population (top).  The bar charts report the frequency with which a synthetic observation in a particular custody level changes to a different custody level and by how many custody levels they would change (e.g., $-2.0$ means the synthetic observation is classified by our random forest model to three custody levels lower than the one the synthetic observation was generated from).}\label{fig:one-col-by-CL-and-race}
\end{figure}

In Figure~\ref{fig:one-col-by-CL-and-race} we see that when we change one column for Black people in our data set who are at custody level 5, there is very large range of custody levels for the resulting synthetic observation whereas for White people for whom we change a single column, the range is much smaller.  We also observe that, unlike in Experiment 3, in this more controlled Experiment 4, we see that people who are identified as Black and in custody levels 1 through 4, are more likely to be kept at the same custody level or to be moved to a higher custody level as a result of changing one column, as compared to similar people who are identified as White.

\subsubsection{Experiment 5} The previous four experiments have perturbed observations to make synthetic observations by sampling from the data that we already have.  In this fifth experiment, we perturb in a different way.  In this experiment we fix all the categorical variables for an observation and generate a new synthetic observation by calculating the margin of error at a 95\% confidence level and then uniformly sampling each quantitative variable from the interval whose radius is equal to the margin of error and whose center is the person's observed value in the data set.  We present this experiment because, while it is trying to capture the same basic idea, it is done in a different way.

\begin{figure}
\centering
\includegraphics[width=\textwidth]{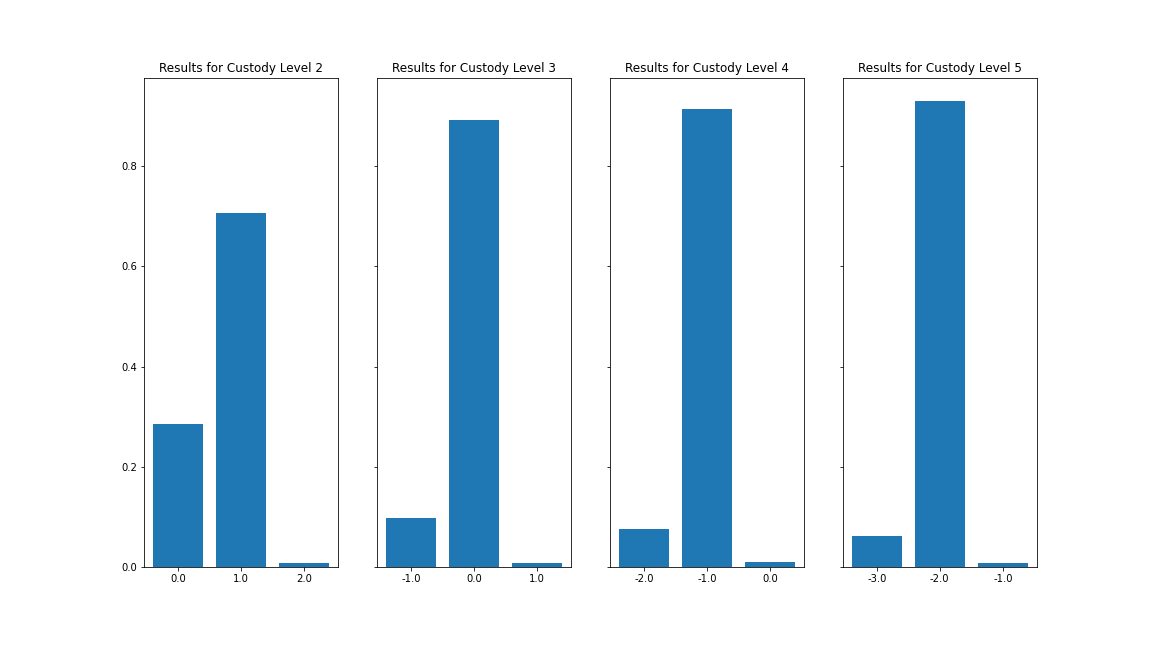}
\caption{Results of Experiment 5 for initial classification.  The bar chart reports the frequency with which a synthetic observation with fixed values for their categorical variables and in a particular custody level changes to a different custody level and by how many custody levels they would change (e.g., $-3.0$ means the synthetic observation is classified by our random forest model to three custody levels lower than the one the synthetic observation was generated from).}\label{fig:95conf-by-CL}
\end{figure}

In Figure~\ref{fig:95conf-by-CL} we observe that the results are rather different and that, in fact, in all but custody level 3, we see that the vast majority of synthetic observations would be classified to a different level than the person from whom they were generated was assigned.  

\subsubsection{Experiment 6}  Our sixth experiment that measures the uncertainty of the PACT tool (defined to be the variability caused by small perturbations to the data) is a sensitivity analysis in which we determine what effect a 10\% increase and decrease to the four most influential variables (namely, the person's prior record score, their offense gravity score, the number of prior commits and their institutional adjustment) has on the predicted custody level.  

\begin{figure}
\centering
\includegraphics[width=\textwidth]{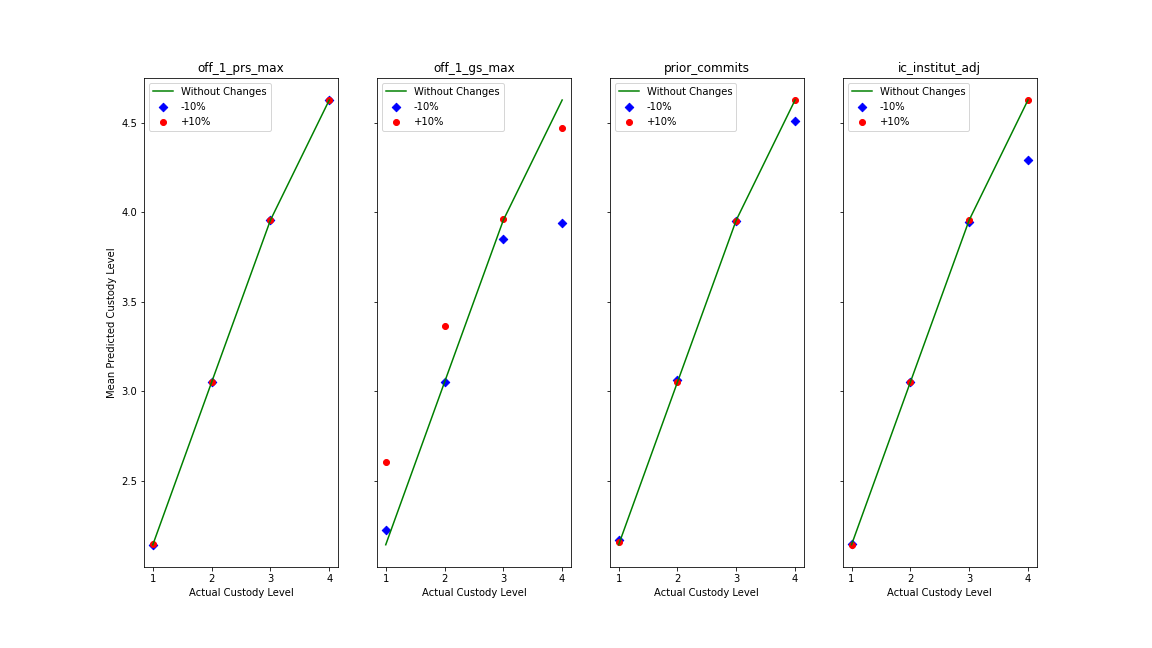}
\caption{Results of Experiment 6 for initial classification.  Each panel reports the average custody level before any changes (green line), after a 10\% increase in the variable corresponding to that panel (a red dot) and after a 10\% decrease in the variable corresponding to that panel (a blue diamond).  Each dot or diamond corresponds to a different initial custody level in the data.}\label{fig:ic_sa}
\end{figure}

In Figure~\ref{fig:ic_sa}, we observe some interesting results.  First, we note that a small change in a person's prior record score has no apparent impact on the average custody level.  Second, we observe that a small change in the number of prior commitments to PADOC a person has also does not have any or much apparent impact on the average custody level.  Third, we see that if a person's custody level is low, a ten percent increase in their associated gravity score increases their custody level and, if their custody level is high, a ten percent decrease lowers their custody level.  Fourth, we see that, for the most part, institution adjustment has little impact on their custody level except in the case when a person was in custody level 5 and their institutional adjustment score is decreased by 10\%.  See Table~\ref{tbl:sa} for more details.

\begin{table}
\begin{tabular}{|l||r|r|r|r|}\hline
Variable change  & CL 2  & CL 3 & CL 4 & CL 5\\\hline
10\% dec. off\_1\_prs\_max & $< 0.1 \%$ & $< 0.1 \%$ & $< 0.1 \%$ & $< 0.1 \%$ \\\hline
10\% inc. off\_1\_prs\_max & $ 0.1 \%$ & $< 0.1 \%$ & $< 0.1 \%$ & $< 0.1 \%$ \\\hline\hline
10\% dec. off\_1\_gs\_max & $3.9 \%$ & $ -0.1 \%$ & $ -2.7 \%$ & $ -13.2 \%$ \\\hline
10\% inc. off\_1\_gs\_max & $22.1 \%$ & $10.3 \%$ & $ 0.2 \%$ & $-3.8 \%$ \\\hline\hline
10\% dec. prior\_commits & $0.8 \%$ & $ 0.4 \%$ & $-0.2 \%$ & $ -3.8 \%$ \\\hline
10\% inc. prior\_commits & $0.7 \%$ & $0.1 \%$ & $ -0.1 \%$ & $-0.4 \%$ \\\hline\hline
10\% dec. ic\_institut\_adj & $0.4 \%$ & $ 0.1 \%$ & $ -0.2 \%$ & $ -7.2 \%$ \\\hline
10\% inc. ic\_institut\_adj & $<0.1 \%$ & $<0.1 \%$ & $ <0.1 \%$ & $<0.1 \%$ \\\hline
\end{tabular}
\caption{We report the relative percent change imparted on custody level by ten percent decreases or increases on the four most important quantitative variables.  An entry in the table that reads $<0.1\%$ means that the relative percent change was negligible;  in particular, that it was between $-0.1\%$ and $0.1\%$.}\label{tbl:sa}
\end{table}

\subsection{Repeated reclassification}

In this section we describe preliminary work in which we attempt to measure and capture the feedback loops that arise with the periodic reclassification that the PADOC carries out.  In this simulation, we sample a fixed number of people in each custody level  and apply the reclassification algorithm and then update data related to the passage of time.  Since we do not have annual data for any individual person, we increment a person's age and update their previous custody level to be the prediction of the previous application of our random forest model.  We fix all the other variables and in this way isolate the effects of time and previous custody level.

We carry out these simulations to get a glimpse of how repeated reclassification works and the impact that it has on individuals and groups (groups of people in the same custody level and groups of people given the same racial identity).  We acknowledge the limitations here that arise from the nature of our data, but believe that this could be a first step in empirically studying the feedback loops that have been studied theoretically in \cite{ensign}.

\begin{figure}
\centering
\includegraphics[width=\textwidth]{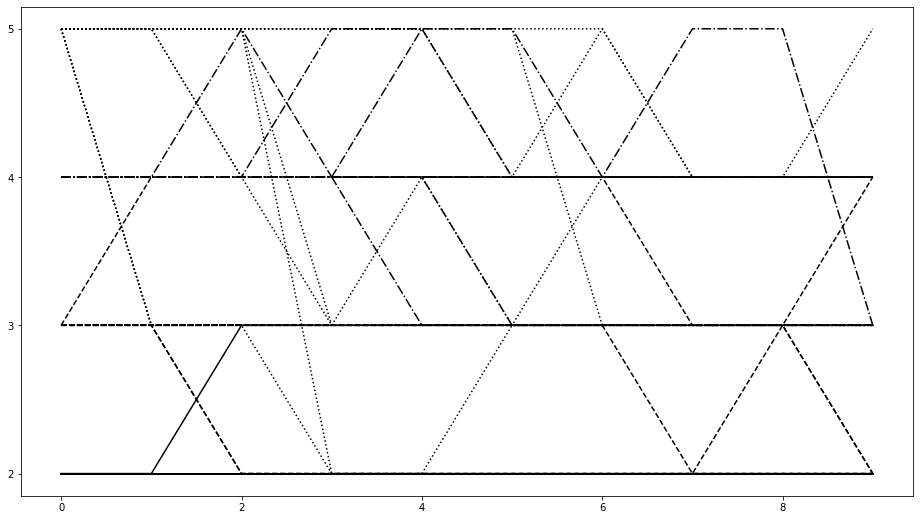}
\caption{The trajectories of 10 individuals at each of the four custody levels.  We repeatedly apply the random forest model for reclassification, updating the person's age and previous custody level.  Each color represents a different person.  While it is hard to follow any individual, it is easy to see that the process is somewhat volatile at the level of individuals.}\label{fig:repeated-10ppl}
\end{figure}

In Figure~\ref{fig:repeated-10ppl} we see the trajectories of 10 individuals starting at each of the four custody levels.  This means that we have 10 individuals in, say, custody level 4 and we apply the reclassification model to each of them.  If the model predicts they change custody level, we update their custody level and increment their age and, if not, we only update their age.  We repeat this process 8 more times and make a plot of the resulting time series.  In this particular image we see, for example, that a person who started at custody level 5 is quickly dropped to custody level 3 and then alternates between levels 3 and 4 for the rest of the time.

\begin{figure}
\centering
\includegraphics[width=.49\textwidth]{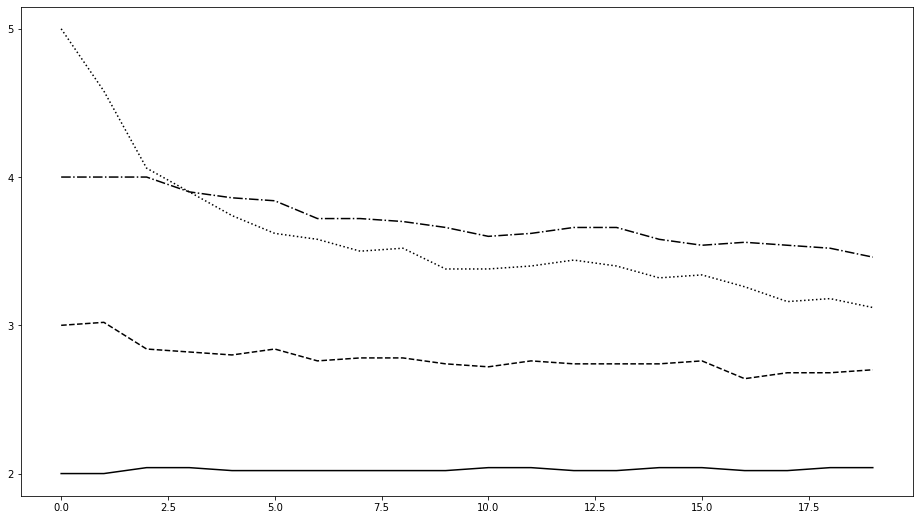}
\hfill
\includegraphics[width=.49\textwidth]{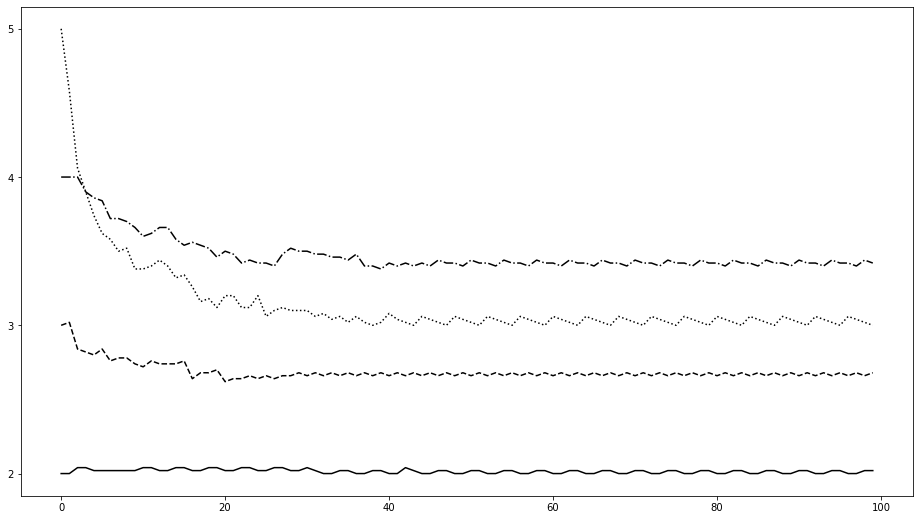}
\caption{The average trajectories of 50 individuals at each of the four custody levels.  We repeatedly apply the random forest model for reclassification, updating the person's age and previous custody level and average across the 50 people who started at each custody level.  We present both the initial behavior over the first 20 years (left) and the long term behavior (right).}\label{fig:repeated-avg}
\end{figure}

Due to the volatility in Figure~\ref{fig:repeated-10ppl}, we also produced Figure~\ref{fig:repeated-avg} in order to understand the aggregate behavior of people who start in each custody level.  We know that as people get older, they tend to be classified at a lower level and so it is not surprising that there is a downward trend for the groups other than those who started at custody level 2.  What is a little surprising is that in the long run, individuals who start at custody level 5, on average, end up at a lower custody level than those who start at custody level 4.  This may be due to how custody level 5 is used by the PADOC and the data we have received (e.g., maybe someone at custody level 5 in our data set is someone who happened to be in solitary confinement when our data was being collected but otherwise is in a lower custody level), but this would require further understanding of how PADOC uses custody levels.

\subsubsection{Differences by race}

\begin{figure}
\centering
\includegraphics[width=.49\textwidth]{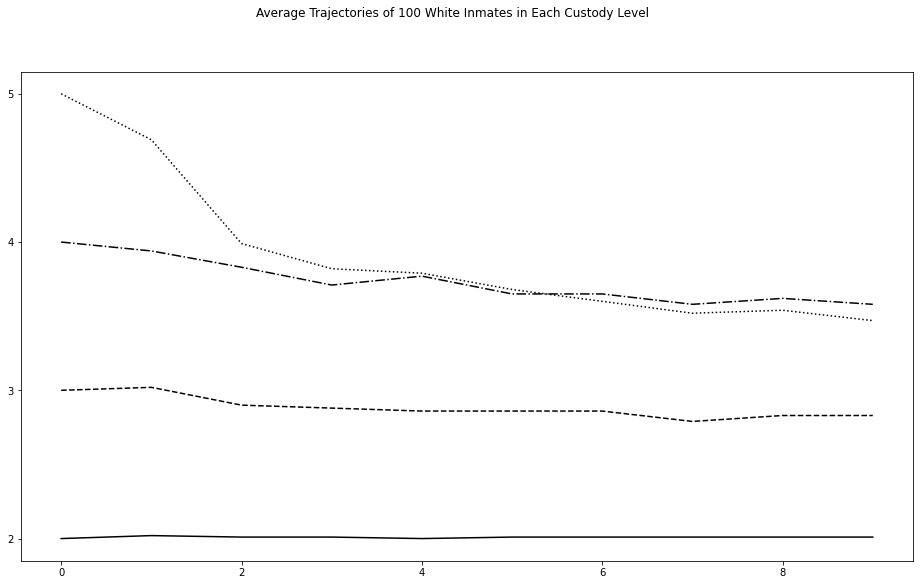}
\hfill
\includegraphics[width=.49\textwidth]{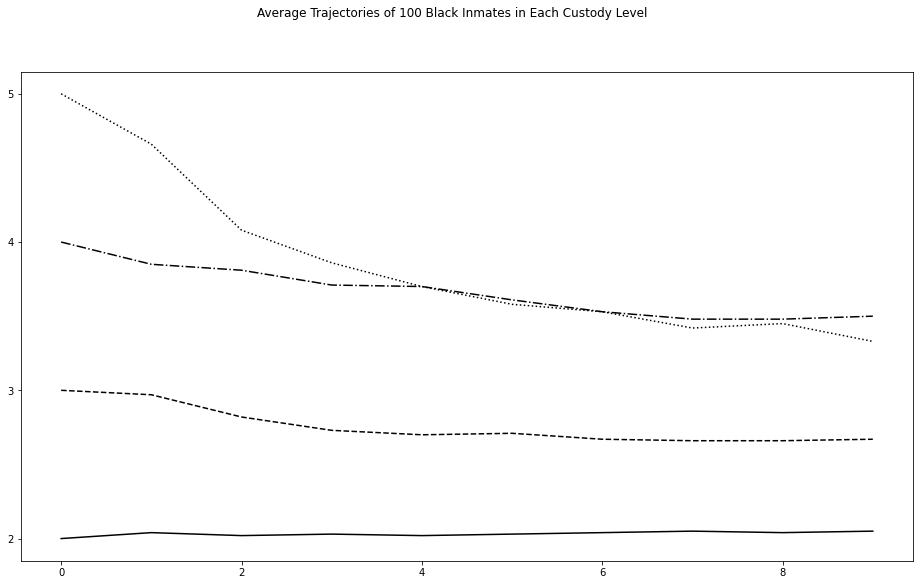}
\caption{The average trajectories for 100 people starting at each custody level and applying the reclassification model 8 times.  We separate them by Black and White people for the sake of comparison.}\label{fig:repeated-B-vs-W}
\end{figure}

In order to study differences in these outcomes by race, we made Figure~\ref{fig:repeated-B-vs-W} in which we take a sample of 100 Black inmates and 100 White inmates, reclassify and then find the average custody level for each sample.  In addition to this visualization, we also decided to measure the variability of a person's trajectory in the following way.  Every time we reclassify a person, we keep a running tally of the absolute value of how many custody levels they change year to year.  So, for example, someone changing from level 3 to 4 would add 1 to the running total and someone changing from 4 to 2 would add 2 to the running total.  For each individual, we calculate these totals and then take the average change per year and for each person in the simulation.  These numbers are reported in Table~\ref{tbl:repeated-B-vs-W}.  

In Figure~\ref{fig:repeated-B-vs-W} and in Table~\ref{tbl:repeated-B-vs-W} we observe two things.  First, they both have the overall downward trajectory that we expect because as people age they tend to be classified to a lower custody level.  Second, Black people have more volatility in their trajectories.  For instance, Black people who start a custody level 2, change custody levels 10 times faster than White people who start at custody level 2.  Black people who start at custody level 4 tend to change custody levels 50\% faster than White people who start at custody level 4.

\begin{table}
\begin{tabular}{|l||r|r|}\hline
Initial custody level & Black people & White people\\\hline
2 & 0.034 & 0.003 \\\hline
3 & 0.087 & 0.099 \\\hline
4 & 0.135 & 0.095 \\\hline
5 & 0.247 & 0.220 \\\hline
\end{tabular}
\caption{Average weighted number of changes in custody level per person, per year for 100 people starting at each custody level and after applying the reclassification 8 times.  We separated it by Black and White people for the sake of comparison.}\label{tbl:repeated-B-vs-W}.
\end{table}

\section{Fairness}\label{sec:cf}

In this section we discuss the fairness of the data and of the random forest models we have developed.  In particular, we ask about the fairness of the outcome,  the fairness of the predictions, the fairness of the decision to override an initial classification, conditioned on the protected classes of race, age and gender.  We emphasize measures of counterfactual fairness as this is similar in spirit to the simulations we have described above.

\subsection{Data-driven fairness of initial classifications and overrides}

There are many ways to determine the fairness of algorithms like initial classification via PACT and the more subjective process of determining whether or not a override is deemed to be warranted for a particular person's classification.  We start by considering whether or not the two decisions are independent of several protected variables (for us these are age, race, and sex).

\begin{table}
\small
\begin{tabular}{|l|l||r|r|}\hline
 $D$ & $a$  & $P(D=1\mid A=a)$ & $P(D=1\mid A=a^\prime)$ \\\hline
Initial custody level $>3$ & Black & $0.56$ & $0.28$\\\hline
Initial custody level $>3$ & Hispanic & $0.51$ & $0.39$\\\hline
Initial custody level $>3$ & Age $>$ 45 & $0.23$ & $0.42$\\\hline
Initial custody level $>3$ & Female & $0.15$ & $0.41$\\\hline\hline
Institutional adjustment $>2$ & Black & $0.54$ & $0.40$\\\hline
Institutional adjustment $>2$ & Hispanic & $0.52$ & $0.47$\\\hline
Institutional adjustment $>2$ & Age $>$ 45 & $0.25$ & $0.52$\\\hline
Institutional adjustment $>2$ & Female & $0.45$ & $0.48$\\\hline\hline
Override to a higher custody level & Black & $0.54$ & $0.37$\\\hline
Override to a higher custody level & Hispanic & $0.42$ & $0.44$\\\hline
Override to a higher custody level & Age $>$ 45 & $0.51$ & $0.43$\\\hline
Override to a higher custody level & Female & $0.27$ & $0.45$\\\hline
\end{tabular}
\caption{A summary of probabilities of various decisions conditioned on a person belonging to a protected group.  The decision $D=1$ corresponds to a positive answer to the first column and the condition $a^\prime$ is the negation of the condition $a$.}\label{tbl:dd-fairness1}
\end{table}

We observe in Table~\ref{tbl:dd-fairness1} the probabilities of being assigned to a high initial custody level does not seem to be independent from whether or not the person is Black, Hispanic, older than 45 years old or female.  We also observe in the same table that the institutional adjustment score a person receives is not independent from whether they are Black or whether they are older than 45 years old.  Finally, we observe that whether a person gets an override to a higher custody level is not independent from whether they are Black or whether they are female.

\subsection{Fairness of the override process}

In addition to the fairness calculations described above, we also want to use the data to determine whether or not an override to a higher custody level was justified by the person having a higher institutional adjustment score. We do this by comparing the probabilities of being given an override to a higher custody level if people have different values for protected variables but the same values for institutional adjustment. 

\begin{table}
\begin{tabular}{|l||r|r|}\hline
$a$ & $P(D=1\mid A=a,\,B=1)$ & $P(D=1\mid A=a^\prime,\, B=1)$\\\hline
Black & $0.75$ & $0.58$\\\hline
Hispanic & $0.69$ & $0.66$\\\hline
Age $>$ 45 & $0.83$ & $0.65$\\\hline
Female & $0.5$ & $0.67$\\\hline
\end{tabular}
\caption{A summary of the probabilities of getting an override to a higher custody $(D=1$) level conditioned on a person belonging to a protected group ($A=a$) and having a high institutional adjustment score ($B=1$). The condition $a^\prime$ is the negation of the condition $a$.}\label{tbl:dd-fairness2}
\end{table}

In Table~\ref{tbl:dd-fairness2}, we see that whether a person is given an override to higher custody level is not independent from whether or not a person is Black, whether they are older than 45 years old or whether they are female but the decision does appear to be independent from whether or not the person is Hispanic.

\subsection{Algorithmic fairness}

In this section we briefly touch on how fair the random forest model is by calculating the statistical conditional parity and the predictive parity of the model.  In particular, if $\hat{Y}$ is the random forest prediction for a high or low custody level, we calculated $P(\hat{Y}=1\mid A=a)$ and $P(\hat{Y}=1\mid A=a)$ and they agreed with the values of $P(D=1\mid A=a)$ and $P(D=1\mid A=a^\prime)$ in the first four columns of Table~\ref{tbl:dd-fairness1} to two decimal places.  So our model performs as fairly as the model run by the PACT.  This is perhaps not surprising due to the high accuracy of our model.

\subsection{Counterfactual fairness}  There are several notions of counterfactual fairness including some that, for example, make use of causal models (see, for example, \cite{kusner}).  In this paper we consider counterfactual fairness for individuals as described by Sokol, \textit{et al.} in \cite{sokol}.  First, certain variables need to be identified as protected: in our case, we use age (a categorical variable that identifies a person as being young ($<25$ years old), of middle age (between 25 and 25 years old) and as being older ($> 45$ years old)), race (a categorical variable whose values are ``White'', ``Black'' or ``Hispanic'', since we have subsetted the data accordingly) and sex.  Second, we need to identify variables to be used by the classifier.  We develop a simpler model and so only include the following variables:  'off\_1\_prs\_max', 'off\_1\_gs\_max', 'prior\_commits', and 'ic\_instit\_adj'.

Next, a classifier and a notion of distance between two points have to be chosen.  In this case we train a $k$-nearest neighbors classifier (classifying whether some on is in a custody level $>3$ or not) with the distance being the taxicab distance.  In the approach, we start with an individual observation and then conduct a brute force grid search through the space of variables and then classify each point generated in this way.  If the grid point and the point that we started with are classified differently and are ``close'' to each other, the grid point is a counterfactual point.  See Table~\ref{tbl:cf-eg} for some examples.  In a sample of 500 observations in our data, 107 (21.4\%) had counterfactuals with protected variables with values that were close to the observations (taxicab distance $\leq$ 3) but were classified to the opposite custody level.

\begin{table} 
\small 
\begin{tabular}{|l||r|r|r|r|}\hline
Initial observation & Class & Counterfactual observation & Class & Distance\\\hline
$(0, 1, 2, 3, 15, 0, 2)$ & Low & age\_cat: 1 $\to$ 0 & High & $1.0$\\\hline
$(0, 1, 1, 1, 12, 1, 2)$ & High & race: 1 $\to$ 0 & Low & $1.0$\\
& & age\_cat: 1 $\to$ 0 & Low & $1.0$\\\hline
$(1, 2, 2, 0.5, 15, 1, 2)$ & Low & age\_cat: 2 $\to$ 1 & High & $1.0$\\
& & race: 2 $\to$ 1 & High & $1.0$\\
& & gender\_female: 1 $\to$ 0, age\_cat: 2 $\to$ 1 & High & $2.0$\\
& & age\_cat: 2 $\to$ 1, race: 2 $\to$ 1 & High & $2.0$\\
& & age\_cat: 2 $\to$ 1, race: 2 $\to$ 0 & High & $3.0$\\\hline 
\end{tabular}
\caption{Examples of counterfactuals for particular data points in our data set.  The tuple in the first column represents a tuple of variables for the observation for which we are looking for counterfactuals: ('gender\_female', 'age\_cat', 'race', 'off\_1\_prs\_max', 'off\_1\_gs\_max', 'prior\_commits', 'ic\_institut\_adj'), the second column indicates whether the observation was in a Low or High custody level, the third indicates how the counterfactual differs from the initial observation, the fourth column tells us what custody level the counterfactual is in and the final column tells us the distance between the counterfactual and the initial observation.  Sex is coded as 0 for male, and 1 for female; age is coded as 0 for young, 1 for middle age and 2 for older; race is coded as 0 for Black, 1 for Hispanic and 2 for White.}\label{tbl:cf-eg}
\end{table}

\section{Discussion}

Algorithms are playing an ever-increasing role in decision-making, in general, and in criminal justice processes, in particular.  The algorithm that we have analyzed, the PACT, is confidential and so the re-creation of the model that we have carried out is necessary to understand it and to critique it.  Among other things, our findings underscore problems that arise from the lack of transparency.  In an earlier paper \cite{dmmr1}, we note that the PADOC justifies keeping the algorithm confidential because otherwise people would ``game'' the algorithm.  This amounts to an admission that the algorithm is not objective.  Moreover, now that we have a fairly accurate model for the PACT, we can determine in which ways it is not objective and what features most influence the custody level assigned by the PACT.  A true commitment to fairness and justice would include transparency.

The lack of transparency is seen in our analysis:  the most important feature in the random forest model is institutional adjustment.  This is a measure of how well a newly incarcerated person is expected to adjust to the particular prison into which they are being committed.  There is no published description of how this score is calculated and so a confidential algorithm like the PACT is using opaque variables like institutional adjustment.  It is also worth pointing out that other important features (namely, the gravity and prior record scores) are also opaque in the sense that there is no published explanation of how these scores are calculated (here we mean that we know an offense might have a gravity score of 12 but there is no published explanation of why certain crimes have a score of 12 and others have a score of 11 and what a difference of one between two scores means).

In addition to having opaque variables heavily influence the assignment of custody levels, we also find a strong emphasis in the model on variables that are static; that is, on variables that an incarcerated person can do nothing to change while they are incarcerated.  For example, we see that the number of prior commitments is important in the classification model that was trained on people some of whom were committed 20 years ago and some of whom were committed a year ago.  As another example, the gravity score is based on the initial charge of the offense (this is also documented as having issues with racial bias in charging and sentencing \cite{vera}), which may have been decades ago for some people.

Demographic or protected variables also lead to very different outcomes and in this sense the PACT has unfair outcomes.  For example, a Black person is twice as likely to be given a high custody level than a non-Black person\footnote{We point out that the conclusions made from the ratios of the last two columns in Table~\ref{tbl:dd-fairness1} and \ref{tbl:dd-fairness2} are likely more conservative than the reality because the non-Black population also includes the Hispanic population which is also treated unfairly}; a younger person is almost twice as likely to be put in a high custody level compared to an older person; and a person identified as male is almost three times more likely to be placed in high custody level than a person identified as female.  Similar disparities persist even when we control for a person's institutional adjustment score.

Not only is the PACT unfair in the ways described above, we claim that its unfair in other ways, too.  For a fair algorithm, it should be the case that a small change to the input does not change the output very much, especially when those changes are to protected variables.  

In Experiment 1, we observed that if we took the existing data set and mixed up the people at each custody level and then reclassified them using the model we trained on the existing data set, almost everybody would change custody levels.  This suggests a certain amount of instability to the algorithm.  In Experiment 3, we first broke up the data into White people and Black people and then see that White people get sent to all custody levels while Black people are sent exclusively to levels 2 and 3.  This suggests that Black people are assigned too high a custody level because if their associated variables were slightly different, they would be assigned to lower custody levels.  In Experiments 2 and 4 we randomly choose a single variable and change that to another value in the data set.  From Experiment 2, we see that for people in custody level 5 there is a lot of variability of where a slightly different observation could be moved to and, in custody levels 2, 3, and 4, there is a significant number of people for whom a small change would send them to a different custody level.  From Experiment 4, the most controlled of the four experiments so far, since we have stratified by custody level and race, and therefore perhaps the most insightful, we see that most people for the most part stay at the same custody level but we see a lot variability for Black people at custody level 5 and a lot of outliers overall.  In Experiment 5, we see that bigger changes in the quantitative variables tends to push everyone towards custody levels 3 and 4 and, in Experiment 6, we see that changes in gravity score and institutional adjustment appear to have the greatest impact on custody level.

Additionally, in the counterfactual fairness analysis, we see that there are people for whom there exist synthetic counterfactual data points that are very close to the original person's observation but who are classified differently.  The closeness here is just between the person's protected variables and the synthetic observation's protected variables.

While we see uncertainty in the classification process as described above, we see a lot of uncertainty (in fact, volatility) in the reclassification process.  We reiterate that our model for reclassification is not trained on very complete data but because we have enough people who have been reclassified, the data that was used for their most recent reclassification and the number of years since their initial reclassification, we are fairly confident in its results.  What we see is that on an individual level, people are being assigned to different custody levels very often.  This might be an example of what geographers have identified as a forced migration \cite{gill}, and will require further study on our part. Prison movement (even within a prison where a custody level determines where the person is house) is used punitively, emerges in immigration detention centers, and taxes detained people and their families. It contradicts other forms of mobility that make modern, middle-class life as well as the immobility experienced by parolees who must maintain a stable address for long periods of time.
 
The volatility is also different by race.  We see that Black people in custody level 2 change custody level 10 times faster than White people in custody level 2.  We see that Black people in custody level 4, change custody level about 50\% faster than White inmates.  In order to achieve fairness, these numbers should not be as different as they are.

\section{Conclusion}  In this paper we have re-created a confidential carceral algorithm used by the PADOC to determine an incarcerated person's custody level and, therefore, their lived experience.  We identified the most important variables in this model and the fairness of the outcomes both in the data and in the model.  We argued for the need to move away from more conventional fairness metrics and towards ways to measuring the uncertainty in these algorithms.  By conducting a series of experiments in which the input data was perturbed in various ways and the associated outputs were analyzed, we show that the PACT tool is unfair, is volatile and, ultimately, uncertain.

\bibliographystyle{plain}
\bibliography{fairness}   

\end{document}